\documentclass[utf8]{frontiersFPHY} 

\usepackage{microtype,subcaption}
\usepackage{times}
\usepackage{soul}
\usepackage{hyperref}
\usepackage[utf8]{inputenc}
\usepackage{caption}
\usepackage{graphicx}
\usepackage{amsmath}
\usepackage{booktabs}
\usepackage{algorithm}
\usepackage{algorithmic}
\usepackage{enumerate}
\usepackage{multirow}
\usepackage{mathrsfs}
\usepackage{amsthm}
\usepackage{amsfonts}
\usepackage{stfloats}
\usepackage{color}
\usepackage[hyphenbreaks]{breakurl}

\def\keyFont{\fontsize{8}{11}\helveticabold }
\def\firstAuthorLast{Dan Lin {et~al.}} 
\def\Authors{Dan Lin\,$^{1,2}$, Jiajing Wu\,$^{1,2,*}$, Qi Yuan\,$^{1,2}$ and Zibin Zheng\,$^{1,2}$}

\def\Address{$^{1}$School of Data and Computer Science, Sun Yat-sen University, Guangzhou, China \\
	$^{2}$National Engineering Research Center of Digital Life, Sun Yat-sen University, Guangzhou, China }
\def\corrAuthor{Jiajing Wu}

\def\corrEmail{wujiajing@mail.sysu.edu.cn}

\begin{document}
\onecolumn
\firstpage{1}

\title[Temporal Weighted Multidigraph Embedding]{T-EDGE: Temporal WEighted MultiDiGraph Embedding for Ethereum Transaction Network Analysis} 

\author[\firstAuthorLast ]{\Authors} 
\address{} 
\correspondance{} 

\extraAuth{}

\maketitle

\begin{abstract}


Recently, graph embedding techniques have been widely used in the analysis of various networks, but most of the existing embedding methods omit the network dynamics and the multiplicity of edges, so it is difficult to accurately describe the detailed characteristics of the transaction networks. Ethereum is a blockchain-based platform supporting smart contracts. The open nature of blockchain makes the transaction data on Ethereum completely public, and also brings unprecedented opportunities for the transaction network analysis. By taking the realistic rules and features of transaction networks into consideration, we first model the Ethereum transaction network as a Temporal Weighted Multidigraph (TWMDG), where each node is a unique Ethereum account and each edge represents a transaction weighted by amount and assigned with timestamp. Then we define the problem of Temporal Weighted Multidigraph Embedding (T-EDGE) by incorporating both temporal and weighted information of the edges, the purpose being to capture more comprehensive properties of dynamic transaction networks. To evaluate the effectiveness of the proposed embedding method, we conduct experiments of node classification on real-world transaction data collected from Ethereum. Experimental results demonstrate that T-EDGE outperforms baseline embedding methods, indicating that time-dependent walks and multiplicity characteristic of edges are informative and essential for time-sensitive transaction networks.

\section{}

\tiny
 \keyFont{ \section{Keywords:} network embedding, Ethereum, machine learning, temporal network, transaction network} 
\end{abstract}

\section{Introduction}

Network is a kind of data form which is often used to describe the relationship between objects. 
The past decade has witnessed an explosive growth of network data which have been used to present information in various areas such as social networks, biological networks, computer networks, financial transaction networks etc. \cite{Volpp2006Complex}. Analysis of large-scale networks has attracted increasing attention from both academia and industry. 
With the rapid development of machine learning technology, how to analyze the data effectively for large-scale complex networks becomes a hot topic in the field of artificial intelligence. 

Financial transaction network exists widely in real world. However, there are relatively few analytical studies on financial transaction networks because the transaction data are usually private for the sake of security and interest. Fortunately, the recent emergence of blockchain technology makes transaction data mining more feasible and reliable. 
Blockchain is a new technology which is described as a innovative application mode of distributed data storage, peer-to-peer transmission, consensus mechanism, encryption algorithm and other computer technologies in the Internet era~\cite{Swan2015Blockchain, Satoshi2008Bitcoin}. Generally speaking, blockchain is a new distributed ledger and the transaction data is stored on the chain in chronological order. 
Ethereum~\cite{wood2014ethereum} is the largest blockchain platform that supports smart contracts.
The Ethereum system introduces the concept of \textit{account} and allocates storage space for recording account balance, transaction time, codes, etc.
Compared with the traditional database, blockchain technology naturally has the characteristics of traceability, anti-tampering and publicity. The openness of public blockchain provides favorable conditions for transaction data mining~\cite{chenweili2018blockchaindata}.

In fact, cryptocurrency and blockchain are highly coupled since blockchain technology is born from Bitcoin. The study of cryptocurrency transaction network has very high application value and
there are already some studies including graph analysis, price prediction, portfolio management, anti-market manipulation, ponzi scheme detection and so on~\cite{RonQuantitative,portfolio2017jiang, Liang2018Evolutionary,feder2018impact,chen2018understanding,weili2019market,Weili2018Detecting}. 
In 2013, Ron \emph{et al.}~\cite{RonQuantitative} described Bitcoin schemes and investigated a large number of statistical properties of the full Bitcoin transaction network. By analyzing the subgraph of the largest transactions, they revealed several characteristics in bitcoin transaction graph: long chains, fork-merge patterns with self loops, keeping bitcoins in “saving accounts” and binary tree-like distributions.
In 2017, Jiang \emph{et al.}~\cite{portfolio2017jiang} presented a deterministic deep reinforcement learning method for cryptocurrency portfolio management. The trading algorithm takes the historic prices of a set of financial assets as input, and outputs the portfolio weights of the set.
In 2018, Liang \emph{et al.}~\cite{Liang2018Evolutionary} traced the properties over time and understood the dynamics of three representative cryptocurrencies, i.e., Bitcoin, Ethereum and Namecoin, by constructing monthly transaction network.

Since it is extremely time-consuming to process the whole blockchain transaction network, it is necessary to find an effective and efficient way to analyse the Ethereum transaction data. As we know, the performance of machine learning tasks depends to a large extent on the selection of data features, so a key problem is how to reasonably represent the feature information in large-scale transaction networks.
In addition, using machine learning-based algorithm often requires feature information of samples, but the account profiles of the transaction networks is often difficult to obtain. The implicit characteristics of the accounts can be mined by means of graph embedding algorithms.

Graph embedding is an effective method to represent node features in a low dimensional space for network analysis and downstream machine learning tasks~\cite{cai2018comprehensive}.
Graph embedding algorithms can effectively reduce the data dimension of transaction network, and transform the large-scale and sparse high-dimensional one-hot node vectors into the dense low-dimensional node vectors. 
Previous graph embedding researches have been present in domains such as social network, language network, citation network, collaboration network, webpage network, biological network, communication network, traffic network, etc ~\cite{cai2018comprehensive}. This implies existing graph embedding techniques may not suitable for transaction network. Using the traditional network embedding algorithm for transaction network analysis will encounter the following challenges: New transactions are generated over time, but existing methods ignore the multiplity and dynamic of transactions. Random walks in transaction networks are meaningful and sequential, but existing methods based on social networks, like DeepWalk and node2vec do not incorporate temporal information.

Random walk mechanism has been widely proved to be an effective technique to measure local similarity of networks for a variety of domains~\cite{spitzer2013principles}.
Among various graph embedding methods, a series of random walk based approaches have been proposed to learn a mapping function from an original graph to a low dimensional vector space by maximizing the likelihood of co-occurrence of neighbor nodes. 
Traditional graph embedding method DeepWalk \cite{perozzi2014deepwalk} verified through experiments that nodes in the random walk sequence and words in the document all follow the power-law, thus word2vec ~\cite{Mikolov2013Efficient} was applied to learn node representations. Similar to DeepWalk, node2vec~\cite{grover2016node2vec} introduced biased random walks which smoothly searches between breadth-first sampling and depth-first sampling strategies.
Recently, to better extract the temporal information from dynamic networks, 
Nguyen \emph{et al.}~\cite{Nguyen:2018:CDN:3184558.3191526} proposed a general framework called Continuous-Time Dynamic Network Embeddings (CTDNE) to incorporate temporal dependencies into existing random walk based network embedding models. 
However, these previous methods omit the network dynamics and the multiplicity of edges, so it is difficult to accurately describe the detailed characteristics of the transaction networks.

To this end, to capture more comprehensive properties of dynamic transaction networks, we propose a novel framework named \textit{\underline{T}emporal W\underline{E}ighted Multi\underline{D}i\underline{G}raph \underline{E}mbedding}  (T-EDGE) for Ethereum transaction network. 
The main contributions of our paper are as follows:
\begin{itemize}
	\item To the best of our knowledge, this is the {\em first} work to understand Ethereum transaction records via graph embedding, which aims to capture the non-negligible temporal properties and important money-transfer tendencies of time-sensitive transaction networks. 
	\item We propose a novel graph embedding method called Temporal Weighted Multidigraph Embedding (T-EDGE) which incorporates transaction information from both time and amount domains, and experiments on realistic Ethereum data demonstrate its superiority over existing methods.
	\item To evaluate our proposed algorithm, we consider an important and practical machine learning tasks, namely node classification with transaction records of phishing and non-phishing accounts collected from Ethereum. The dataset can be accessed on XBlock (\url{xblock.pro}).
\end{itemize}

The remainder of this paper is organized as follows. 
First, Section~\ref{sec:framework} demonstrates our working flow for Ethereum transaction network analysis. Then, Section~\ref{sec:modeling} describes how we model the transaction records as a temporal weighted multidigraph. 
Then, we introduce our proposed network embedding algorithm T-EDGE in Section~\ref{sec:NetworkEmbedding}, and evaluate our algorithm by conducting node classification in Section \ref{sec:NodeClassification}.
Finally, Section \ref{sec:Conclusion} concludes the paper. 

\section{Related work}

\subsection{Transaction analysis of cryptocurrencies}


Nan et al. [146] proposed a method for bitcoin mixing detection that consisted of different stages: Constructing the Bitcoin transaction graph, implementing node embedding, detecting outliers through AE

\section{Framework}
\label{sec:framework}
\begin{figure*}[htbp]	
	\centering	
	\includegraphics[width=\linewidth]{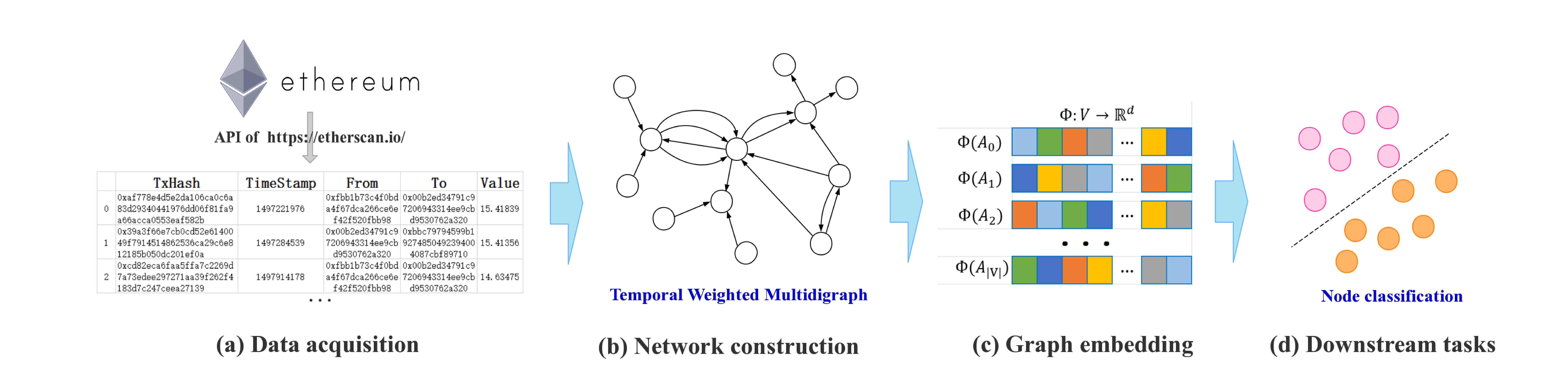}	
	\centering
	\caption{The architecture of the proposed framework for network analysis of Ethereum.}	
	\label{fig:framework}	
\end{figure*}

In this section, we describe the working flow of Ethereum transaction network analysis presented in this work.
As Figure~\ref{fig:framework} shows, we demonstrate the four main steps of the proposed framework for Ethereum transaction network analysis, including data acquisition, network construction, graph embedding and downstream tasks.

\begin{enumerate}[(a)]
	\item \textbf{ Data acquisition.} The data collection is the basis of transaction network analysis. Thanks to the openness of blockchain, researchers are able to autonomously access Ethereum transaction records. Through the API of Etherscan (\url{etherscan.io}), a block explorer and analytics platform for Ethereum, we can easily obtain the historical transaction data of the target account. 
	As the size of the total transaction records is extremely large, we adopt the $K$-order subgraph sampling method~\cite{lin2020modeling}, to obtain the local structure of the target accounts.
	
	\item \textbf{ Network construction.} 
	This step abstracts the original transaction record into a network structure for further analysis.
	In most of existing studies on blockchain transaction networks, the transaction networks are usually constructed into simple graphs, that is, multiple transactions between a pair of accounts are merged into one edge, thus ignoring the multiplicity and dynamics of transactions between accounts. Different from prior work, in this work, we model the multiple interactions between accounts as a Temporal Weighted Multidigraph~\cite{lin2020modeling} to facilitate a more comprehensive analysis of transaction behaviors. 
	
	\item \textbf{ Graph embedding.} In the framework of Ethereum transaction network analysis, the role of network embedding is to mine the implicit features of accounts in the transaction network and reduce the transaction data dimension. In order to learn the meaningful node representation vectors in the dynamic transaction network, we propose the improved embedding algorithm called Temporal Weighted Multidigraph Embedding (T-EDGE) based on temporal random walk. T-EDGE aims to capture the time and amount information that cannot be ignored in Ethereum transaction network.
	
	\item \textbf{ Downstream tasks.} We evaluate our model by conducting experiments on a typical machine learning task, namely node classification. The good performance of the downstream tasks reflects the effectiveness of embedding methods. Besides, analytical applications can be regarded as the ultimate goal of the Ethereum transaction network embedding. In this paper, we incorporate two current hot topics -- cryptocurrency transaction analysis and machine learning, and use machine learning technology to help us make more accurate predictions about the future of Ethereum transaction network.

\end{enumerate}

\section{Ethereum Transaction Network}
\label{sec:modeling}

Being the largest public blockchain-based platform that supports smart contract, Ethereum introduces the concept of \textit{account} to facilitate the implementation of smart contracts.
An Ethereum account is formally an address, but adds storage space for recording account balances, transactions, codes, etc. 
Ethereum addresses are composed of the prefix "0x", a common identifier for hexadecimal, concatenated with the rightmost 20 bytes of the public key. One Example is ``0x00b2ed34791c97206943314ee9cbd9530762a320'' 
The corresponding cryptocurrency on Ethereum, known as \textit{Ether}, can be transferred between accounts and used to compensate participant mining nodes.

Ethereum blockchain consists of infinite linked blocks, which can be viewed as data-packages, including a series of transactions and some other information. 
In details, the transaction data packages obtained from the Etherscan website are as followed: the \textit{TxHash} field is a unique 66 characters identifier of a transaction, the \textit{Value} field is the value being transacted in Ether, and the~\textit{Timestamp} field is the time at which a transaction is mined. Besides, the \textit{From} and \textit{To} field is the sending party and receiving party of a transaction.

\begin{figure*}[htbp]	
	\centering	
	\includegraphics[width=\linewidth]{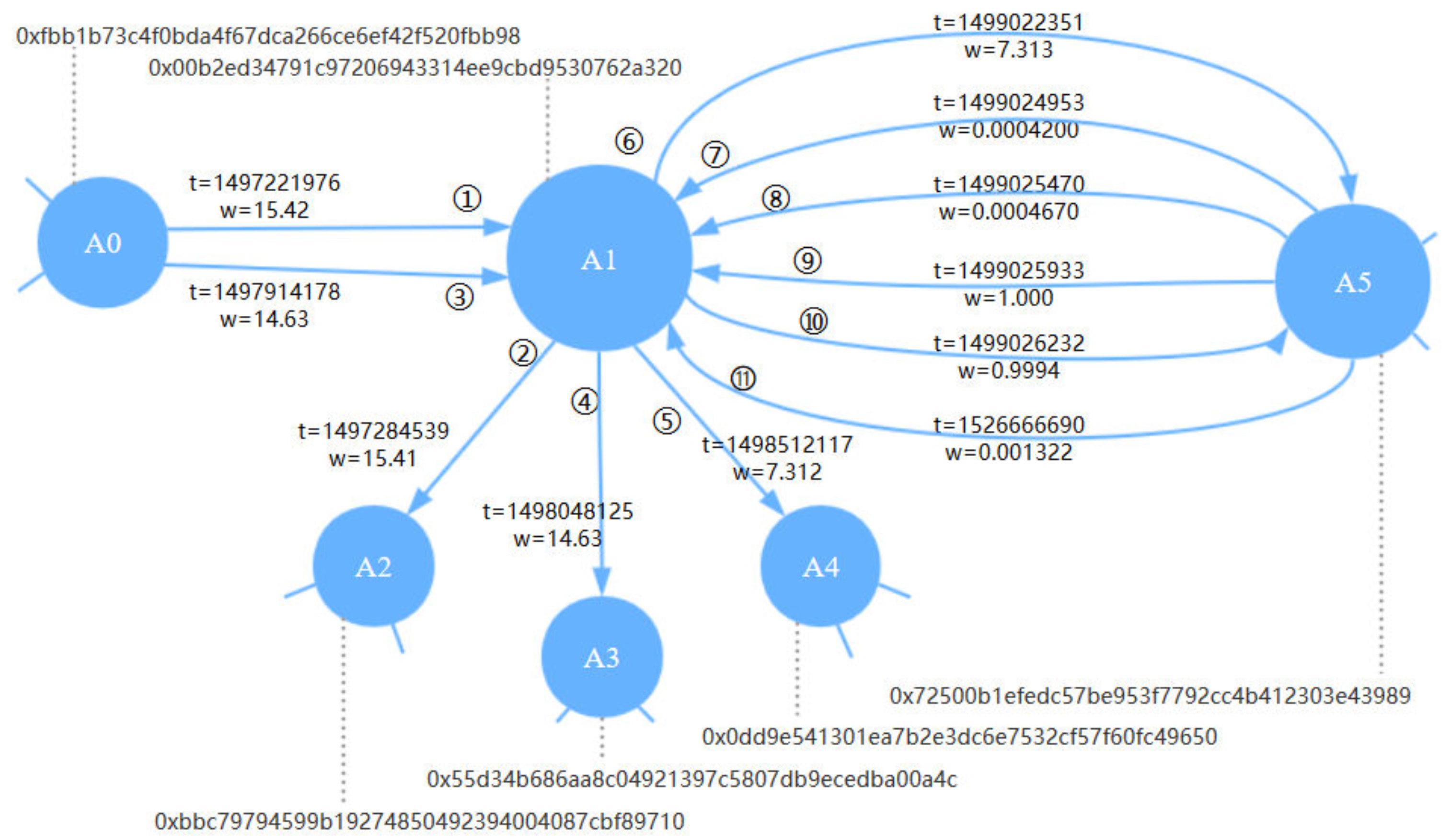}	
	\caption{The schematic of Ethereum transaction network. Each nodes represents an Ethereum account. Each edge represents a transaction, attached by timestamp $t$ and amount value $w$ (in \textit{Ether}), and indexed in the increasing order of $t$.} 	
	\label{fig:twmdg}	
\end{figure*}

In this section, we abstract the original transaction record as a Temporal Weighted Multidigraph (TWMDG).
Figure~\ref{fig:twmdg} is a microcosm of transaction activities on Ethereum. 
In prior work on blockchain transaction network analysis, the transaction network was constructed as a simple network, that is, multiple transactions between nodes were accumulated as one edge. The multiplicity and dynamics of transactions between accounts were ignored. 
Therefore, we adopt Temporal Weighted Multidigraph (TWMDG) to represent Ether transfer between accounts more comprehensively.

Based on collected four-tuples \textit{(From, To, Value, Timestamp)}, we can model the Ethereum transaction records as a Temporal Weighted Multidigraph $G=(V,E)$, where each node represents a unique account and each edge represents a unique Ether transfer transaction. In such a graph, $V$ is the set of nodes and $E$ is the set of edges. Each edge is unique and is represented as $e = (u, v, w, t)$, where $u$ is the source node, $v$ is the target node, $w$ is the weight value and $t$ is the timestamp.

\section{Network embedding}
\label{sec:NetworkEmbedding}

In the analysis of Ethereum transaction network, our goal is to learn an embedding vector for each node, the purpose being to mining the implicit characteristics of nodes in the transaction network and incorporating the time and amount information of the transaction network into the node vector.
For the network model TWMDG built in the previous section, this paper proposes an improved network embedding algorithm based on a random walk. 
We now define specific problem as follows.

\textbf{\underline{T}emporal W\underline{E}ighted Multi\underline{D}i\underline{G}raph \underline{E}mbedding (T-EDGE)}: Given a temporal weighted multidigraph $G=(V,E)$, let $V$ be the set of nodes and $E$ be the set of edges. Each edge is unique and is represented as $e = (u, v, t, w)$, where $u$ is the source node, $v$ is the target node, $t$ is the timestamp and $w$ is the weight (Specifically, $w$ is the amount value of transaction in Ethereum transaction network). We define following mapping functions: For $\forall e\in E$, $Src(e)=u$, $Dst(e)=v$. Function $W(e)=w$ maps an edge to its weight and function $T(e)=t$ maps an edge to its timestamp. Our principal goal is to learn an embedding function $\Phi:V\rightarrow \mathbb{R}^d$ ($d\ll|V|$) which preserves original network information.



The learned representations aim to include node similarity, as well as temporal and weighting properties specifically for financial transaction networks, thus enhancing predictive performance on down-stream machine learning tasks. The proposed method T-EDGE learns more appropriate and meaningful dynamic node representations using a general embedding framework consisting of two main parts: 
\begin{itemize}
	\item The first part is the temporal walk generator with the temporal restriction and walking strategies.
	\item The second part is the update process based on skip-gram, and the parameters are updated by Stochastic Gradient Descent algorithm. \newline 
\end{itemize}

\subsection{Random Walk}

For scalable network representation learning, random walk mechanism has been widely proven to be an effective technique to capture structural relationships between nodes~\cite{perozzi2014deepwalk}.
We employ a temporal walk for transaction networks by considering temporal dependencies and multiplicity of edges. This kind of random walk sequences contains the practical meaning of money flow in transaction networks.

In a temporal weighted multidigraph, the \textit{temporal walk} defined as sequential incremental path from the beginning node to the end node. Such a temporal walk is represented as a sequence of $l$ nodes $ walk_n=\{v_1, v_2, ..., v_l\}$ together with a sequence of $(l-1)$ edges $ walk_e=\{e_1, e_2, ..., e_{l-1}\}$, where $Src(e_i)=v_i$, $Dst(e_i)=v_{i+1}$ $(1\leq i \leq(l-1))$, and $T(e_i)\leq T(e_{i+1})$ $(1\leq i \leq (l-2))$. This temporal restriction is a novel idea designed for the temporal walk.

Consider a temporal walk that just traversed edge $e_{i-1}$, and is now stopping at node $v_i$ at time $t=T(e_{i-1})$. The next node $v_{i+1}$ of the random walk is decided by selecting a temporally valid edge $e_i$. 
We define the \textit{temporal edge neighborhood} for a node $u$ as $N_t(u) = \{ e~|~Src(e)=u, T(e)\geq t \} $. Let $\eta_-: \mathbb{R}\rightarrow \mathbb{Z}^+$ to be a function that maps the timestamps of edges to a descending ranking, and let $\eta_+:\mathbb{R}\rightarrow \mathbb{Z}^+$ be a function that maps the timestamps of edges to an ascending ranking. Here are our walking strategies used in Ethereum transaction network embedding:

\textbf{T-EDGE}. 
In our temporal weighted multidigraphs discussed here, a random walk generator of T-EDGE samples uniformly from the neighbors. All candidate edges in $N_t$ have the same probability to be selected as the next edge of random walk. The expression of the probability is
\begin{equation}
\label{equ:time_uniform}
P(e)=\frac{1}{|N_t(v_i)|}.
\end{equation}

\textbf{T-EDGE (TBS)}. 
TBS refers to Temporal Biased Sampling. For financial transaction networks, the similarity between accounts is time-dependent and dynamic. Naturally, there is a strong transaction relationship between accounts with frequent transactions.
The probability of selecting each edge $e\in N_t(v_i)$ can be given as:
\begin{equation}
\label{equ:time_close_linear}
P(e)=P_{TBS}(e)=\frac{\eta_-(T(e))}{\sum_{e'\in N_t(v_i)}~\eta_-(T(e'))}.
\end{equation}	

\textbf{T-EDGE (WBS)}. 
WBS refers to Weighted Biased Sampling. The weight value of each transaction indicates the significance of interactions between the two accounts involved. 
The transaction amount can reflect the importance of transactions between accounts and then reflect the degree of correlation between accounts. In most cases, there is a strong similarity between accounts with large amount of transactions.
The probability of each edge $e\in N_t(v_i)$ being selected is

\begin{equation}
\label{equ:amount_linear}
P(e)=P_{WBS}(e)=\frac{\eta_+(W(e))}{\sum_{e'\in N_t(v_i)} \eta_+(W(e'))}.
\end{equation}

\textbf{T-EDGE (TBS+WBS)}. We combine the aforementioned sampling probabilities considering information from both temporal and weighted domains by 
\begin{equation}
\label{equ:TBS+WBS}
P_{TBS+WBS}(e) = P_{TBS}(e)^\alpha P_{WBS}(e)^{(1-\alpha)} , (0\leq\alpha\leq1),
\end{equation}
\begin{equation}
\label{equ:TBS+WBS_2}
P(e)=\frac{P_{TBS+WBS}(e)}{\sum_{e'\in N_t(v_i)} P_{TBS+WBS}(e')},
\end{equation}	

for  $\forall e\in N_t(v_i)$. Here $\alpha=0.5$ is the default value for balancing between TBS (time domain) and WBS (amount domain).\newline 

\begin{table}[tbp]
	\centering
	\begin{tabular}{ccccc}
		\toprule
		
		\multirow{2}[4]{*}{\textbf{Algorithms}} & \multicolumn{2}{c}{\textbf{Time domain}} & \multicolumn{2}{c}{\textbf{Amount domain}} \\
		\cmidrule{2-5}    \multicolumn{1}{c}{} & \multicolumn{1}{c}{ Unbiased } & Biased & \multicolumn{1}{c}{ Unbiased } & Biased   \\
		\midrule
		T-EDGE & \multicolumn{1}{c}{$\surd$} & \multicolumn{1}{c}{} & \multicolumn{1}{c}{$\surd$} & \multicolumn{1}{c}{} \\
		\midrule
		T-EDGE (TBS) &       & $\surd$ & \multicolumn{1}{c}{$\surd$} & \multicolumn{1}{c}{} \\
		\midrule
		T-EDGE (WBS) & \multicolumn{1}{c}{$\surd$} & \multicolumn{1}{c}{} &       & $\surd$ \\
		\midrule
		T-EDGE (TBS+WBS) &       & $\surd$ &       & $\surd$ \\
		\bottomrule
	\end{tabular}%
	\caption{Four types of T-EDGE variation for Ethereum transaction network.}
	\label{tab:table1}%
\end{table}%

When ending up with a leaf node, we return the walk immediately. This setting is just the same as the compared methods, DeepWalk and node2vec.

Note that T-EDGE can be regarded as a specific version of DeepWalk for temporal and directed multigraphs like the transaction networks. 
As Table~\ref{tab:table1} shows, all candidate edges (temporal edge neighborhood) are equally likely to be selected by T-EDGE.
T-EDGE (TBS), T-EDGE (WBS) denote to add sampling preference on the time domain and the amount domain respectively. T-EDGE (TBS+WBS) means to add sampling preference on both the time domain and the amount domain. \newline


\subsection{Learning Process}

In the previous subsection, we described how to get the sampling sequence of temporal walk related to time and weight. In this part, we will formally describe the process of learning node vectors using skip-gram model~\cite{Mikolov2013Efficient,mikolov2013distributed}. 


\begin{table}[tbp]
	\centering
	
	\begin{tabular}{ccccc}
		\toprule
		\textbf{Research domain} & \textbf{Example} &  \textbf{Input} & \textbf{Output} \\
		\midrule
		Natural language processing & word2vec & Sequence of word (sentences) & Word vectors \\
		\midrule
		Graph representation learning & deepwalk & Sequence of nodes (random walks) & Node vectors \\
		\bottomrule
	\end{tabular}%
	\caption{The comparison between language model word2vec and graph model Deepwalk.}
	\label{tab:analogy}%
\end{table}%

The essence of skip-gram model is a three-layer neural network model, including input layer, hidden layer and output layer. First, we train a neural network model based on the sampling walk sequences, but the purpose of training is not to use the model to predict the test set, but to use the parameters learned from the model, namely the hidden layer parameters, as our node vectors. Then, by making an analogy between natural language sentence and truncated random walk sequence (as shown in Table~\ref{tab:analogy}), node representations are then learned by maximizing the probability of observing the neighborhood of a node conditioned on its embedding. This cost function is as followed:
\begin{equation}
\label{equ:objective}
\min_{\Phi} -\Pr(\{v_{i-k}, ..., v_{i+k}\} \backslash v_i | \Phi(v_i)),
\end{equation}
where $k$ is the window size. 
According to the conditional independent assumption in skip-gram model, we have:
\begin{equation}
\label{equ:independent}
\Pr(\{v_{i-k}, ..., v_{i+k}\} \backslash v_i | \Phi(v_i)) = \prod_{j=i-k, \newline j\neq i}^{i+k}~\Pr(v_j|\Phi(v_i)).
\end{equation}

Similar to DeepWalk, we employ ``hierarchical softmax" technique~\cite{perozzi2014deepwalk} to accelerate the computation of  $\Pr(v_j|\Phi(v_i))$. We first apportion $|V|$ nodes to the leaf nodes of a Huffman Tree, and then transform the computation of $\Pr(v_j|\Phi(v_i))$ into computing the probability of walking randomly from the root of Huffman Tree with inputting node $v_i$ and outputting node $v_j$. The probability is 
\begin{equation}
\label{equ:h-softmax}
\Pr(v_j | \Phi(v_i)) = \prod_{t=1}^{\lceil \log |V| \rceil}~\Pr(b_t|\Phi(v_i)),
\end{equation}
where $b_t$ is from $\{ b_0=root, b_1, ..., b_{\lceil \log |V| \rceil}=v_j\}$. Then we model $\Pr(b_t|\Phi(v_i)$ with $sigmiod$ function:
\begin{equation}
\label{equ:h-softmax-sigmiod}
\Pr(b_t|\Phi(v_i)=\frac{1}{1+\exp(-\Phi(v_i)\cdot \Phi(b_{t-1}))},
\end{equation}
where $\Phi(b_{t-1})$ is the representation of $b_t$'s parent node in the Huffman tree. Skip-gram model then uses back propagation algorithm and Stochastic Gradient Descent to update the weight. 

Random walk based graph embedding methods have been proved to be scalable and effective for large graphs. The time complexity of the temporal walk part and the skip-gram learning procedure is $O(r|V|L)$ and $O(|V| \log |V|)$, respectively, where $|V|$ is the number of nodes, $r$ denotes walks per node, and $L$ refers to the length of random walk.

\section{Experiments and Results}
\label{sec:NodeClassification}

Downstream tasks such as node classification are commonly considered for the verification of graph embedding methods. To evaluate the performance of the proposed T-EDGE algorithms, we conduct node classification experiments to classify the labeled phishing accounts and unlabeled accounts (treated as non-phishing accounts) on Ethereum. The better performance of classification demonstrates that our T-EDGE algorithms outperforms baseline embedding methods, on the same time, node classification for detecting phishing accounts on Ethereum is also of great value. Phishing scam is a new type of cybercrime which arises along with the emergence of online business~\cite{Liu2001Introduction}. It is reported to accounts for more than 50\% of all cyber-crimes in Ethereum since 2017~\cite{konradt2016phishing}. \newline

\subsection{Data Acquisition}
\label{sec:DataAcquisition}

To train our node classification model using supervised learning, we obtain 445 phishing nodes labeled by Etherscan, and the same number of randomly selected unlabeled nodes as our objective nodes.

\begin{figure}[htb]
	\centering	
	\includegraphics[width=110mm]{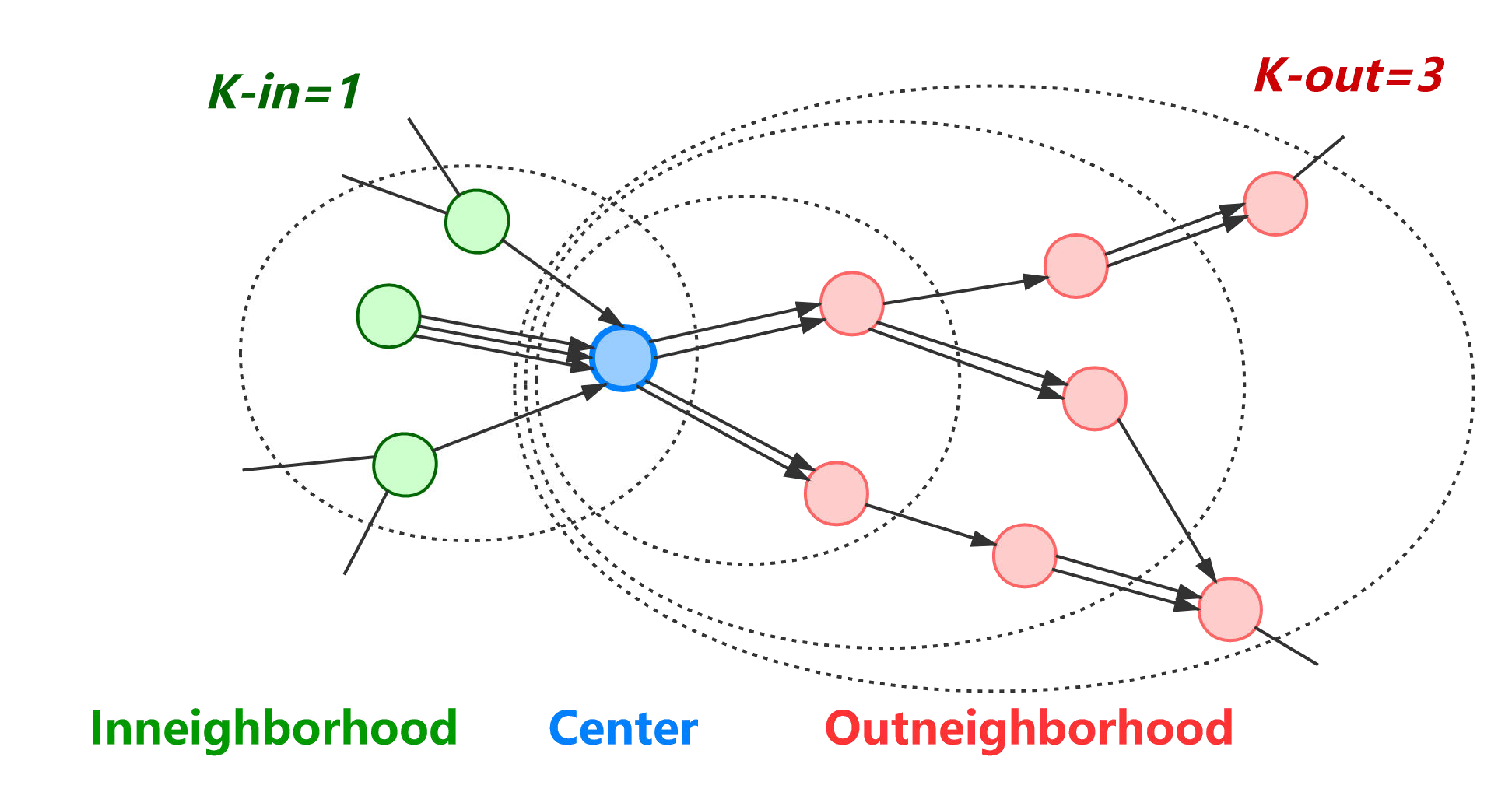}
	\caption{A schematic illustration of a directed $K$-order subgraph for phishing node classification.}	
	\label{fig:korder}	
\end{figure}

$K$-order sampling is an effective method to obtain the local information of an objective accounts~\cite{lin2020modeling}.
Centered by each objective account, we obtain a directed $K$-order subgraph, where $K$-in and $K$-out are two parameters to control the depth of sampling inward and outward from the center, respectively.
As shown in Fig.~\ref{fig:korder}, we make an assumption that for a typical \textit{Ether} transfer flow centered on a phishing node, the previous node of the phishing node may be a victim, and the next one to three nodes may be the bridge nodes with money laundering behaviors. Therefore, we collect subgraphs with $K$-in = 1, $K$-out = 3 for each of the 890 objective nodes and then splice them into a large-scale network with 86,623 nodes. \newline

\subsection{Setting} In the experiments, we compare the proposed T-EDGE algorithms with two baseline random-walk based graph embedding methods: 

\begin{itemize}
	\item \textbf{DeepWalk} is the pioneer work in employing random walks to learn a latent space representation of social interactions. Borrowing the idea of word2vec, the learned representation encodes community structure so it can be easily exploited by standard classification methods~\cite{perozzi2014deepwalk}.
	\item \textbf{Node2vec} further exploits a flexible neighborhood sampling strategy, i.e., Breadth-first Sampling (BFS) and Depth-first Sampling (DFS), with parameters $p$ and $q$ to capture both local and global structure~\cite{grover2016node2vec}. 

\end{itemize}

To ensure a fair comparison, we implement the directed version of DeepWalk and node2vec using OpenNE (an open-source package for network embedding, \url{github.com/thunlp/openne}). For these random walk based embedding methods, we set several hyperparameters: the node embedding dimension $d=128$, the size of window $k=4$, the length of walk $l=10$, and walks per node $r=4$.
For node2vec, we grid search over $p,q\in \{ 0.50, 1.0, 1.5, 2.0\}$ according to \cite{grover2016node2vec}. For DeepWalk, we set $p=q=1.0$ as it is a special case of node2vec.
We implement the Skip-gram model by using a Python library named Gensim~\cite{rehurek2010}, a framework for fast Vector Space Modelling.
 \newline

\subsection{Metrics}

To make a comprehensive evaluation, we randomly select \{50\%, 60\%, 70\%, 80\%\} of objective nodes as training set and the remaining objective nodes as test set respectively. 
We train a classic binary classifier, namely, Support Vector Machine (SVM) with the training set, to classify the samples of the test set.
Note that we use five-fold cross validation to train the classifier and evaluate it on the test set. 

For a binary classification task based on a supervised learning framework, it can be divided into the following four cases according to the actual labels of the samples and the prediction results of the classifier.

\begin{itemize}
	\item True Positive (TP): Samples whose labels are positive and also predicted to be positive.
	\item True Negative (TN): Samples whose labels are positive but predicted to be negative.
	\item False Positive (FP): Samples whose labels are negative but predicted to be positive.
	\item False Negative (FN): Samples whose labels are negative and also predicted to be positive.	
\end{itemize}

In classification tasks, micro-F1 (Mi-F1) and macro-F1 (Ma-F1) are generally used to evaluate classification accuracy. First we have
\begin{itemize}
	\item \textit{precision}: $ \frac{TP}{TP+FP},$
	\item \textit{recall}: $ \frac{TP}{TP+FN}.$
\end{itemize}

F1-score is an indicator used to measure the accuracy of the binary classification model. 
The calculation formula is

\begin{equation}
\label{equ:f1-score}
 2 \times \frac{precision\times recall }{precision + recall}.
\end{equation}
Macro-F1 refers to calculating the total $precision$ and $recall$ of all categories for F1-score, while Micro-F1 refers to the calculation of F1-score after calculating the average of $precision$ and $recall$ for each category. 
\newline

\subsection{Reults}

\begin{figure*}[tbp]	
	\centering	
	\includegraphics[width=\linewidth]{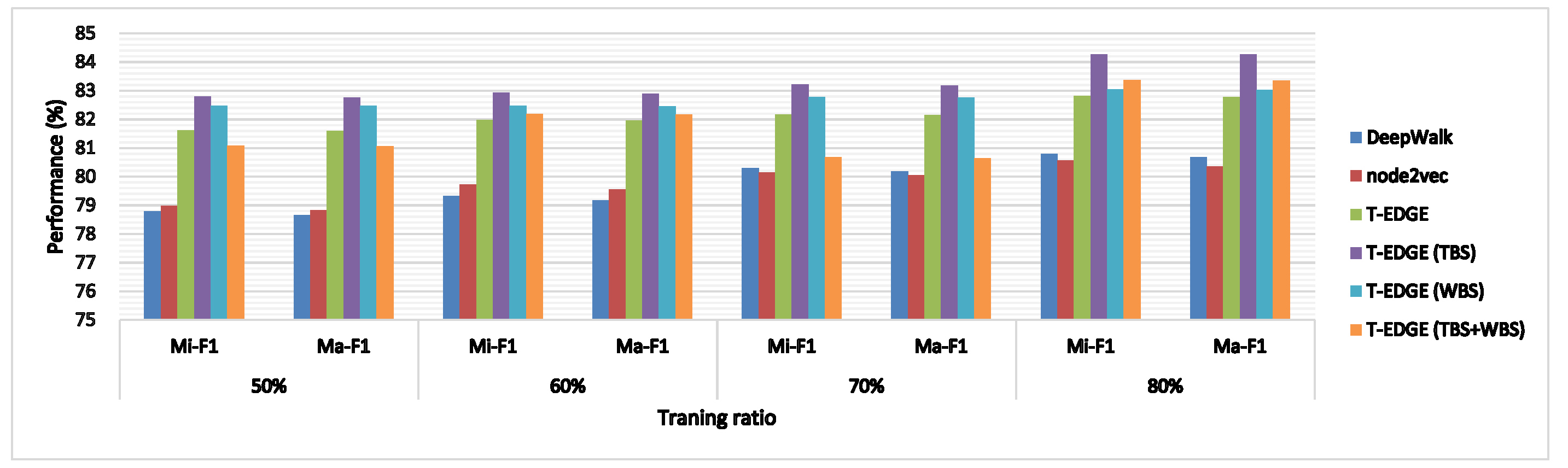}
	\caption{Node classification performance with different training ratio.}	
	\label{fig:result1}	
\end{figure*}

The results of micro-F1 (Mi-F1) and Marco-F1 (Ma-F1) are shown in Fig.~\ref{fig:result1}.
According to  Fig.~\ref{fig:result1}, we have the following observations: 

\begin{enumerate}[(1)]
	\item  Our proposed methods {T-EDGE, T-EDGE (TBS), T-EDGE (WBS), T-EDGE (TBS+WBS)} overwhelmingly outperform DeepWalk and node2vec;
	
	\item  Both T-EDGE (TBS) and T-EDGE (WBS) attain better performance than T-EDGE in which the random walk generator has uniform probability;
	
	\item  Both T-EDGE (TBS) and T-EDGE (WBS) perform better than T-EDGE (TBS+WBS) which considers both temporal and amount information with parameter $\alpha=0.5$.
		
\end{enumerate}

All in all, our proposed methods learn effective node representations incorporating rich information, which does help us get better performance in classification task. The result also indicates that time-dependent walks and edge information are essential in transaction networks. \newline

\subsection{Parameter analysis of $\alpha$}

\begin{figure}[hbtp]	
	\centering	
	\includegraphics[width=110mm]{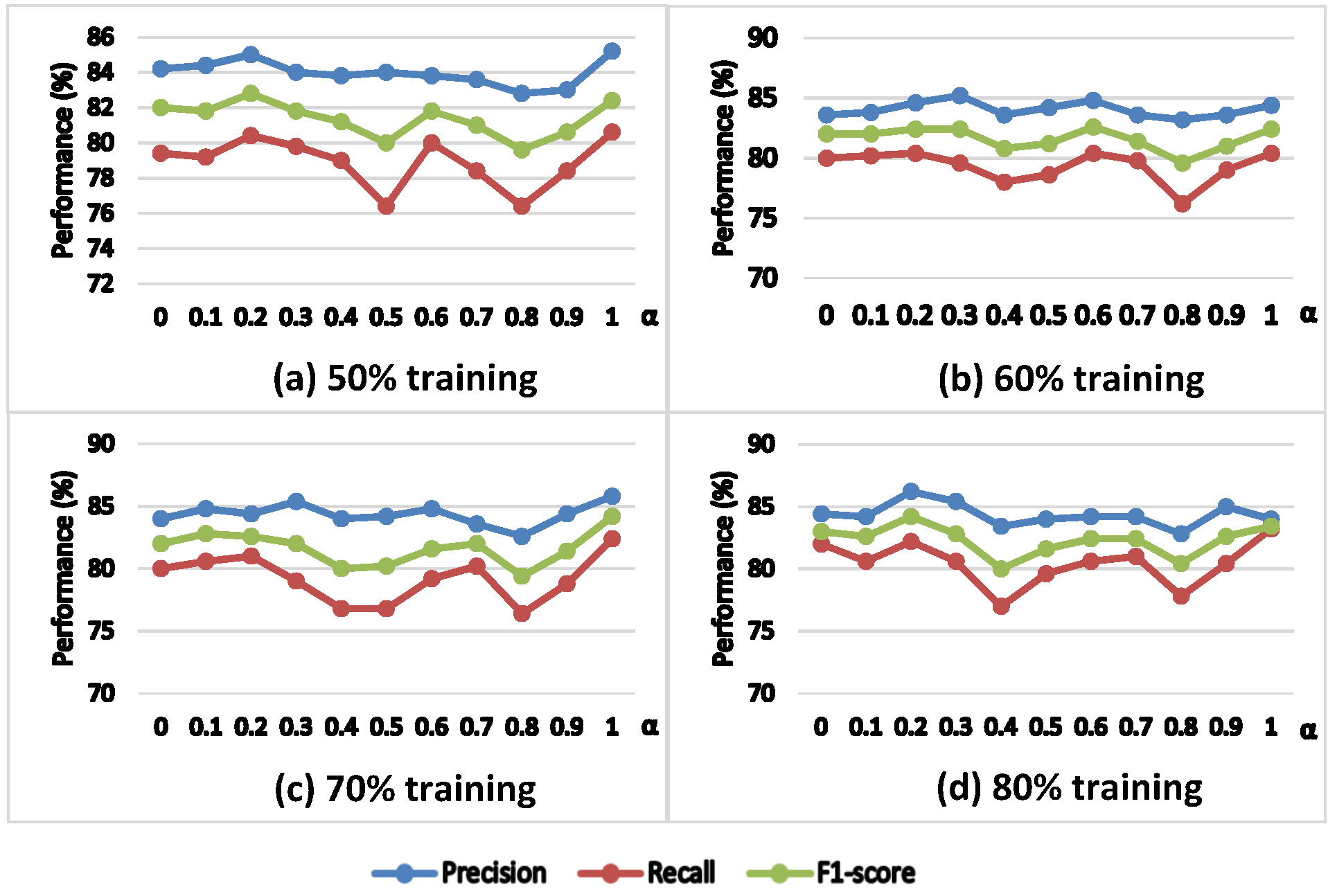}
	\caption{The classification performance on the parameter $\alpha$ with different training ratio. }	
	\label{fig:alpha}	
\end{figure}

Furthermore, the third observation mentioned above inspires us to analyze the coupling parameter $\alpha$. Larger $\alpha$ means considering more time domain information in random walk, while smaller $\alpha$ means more amount domain information.
Fig.~\ref{fig:alpha} compares the classification performance on the parameter $\alpha$ with different training ratio, in terms of $precision$, $recall$ and F1-score. We find that $\alpha=0.8$ is a poor choice but there is no single  $\alpha$ that is a clear winner. Nevertheless, we can observe that  $\alpha \in [0.2,0.3]$ and $\alpha=1$ are relatively better choices. This result indicates that it is better to consider or favor a single strategy or favor a single strategy than to consider both strategies equally at the same time.

\section{Conclusion}
\label{sec:Conclusion}

In this work, we proposed a novel framework for Ethereum analysis via network embedding. Particularly, we constructed a temporal weighted multidigraph to retain information as much as possible and present a graph embedding method called T-EDGE which incorporates temporal and weighted information of financial transaction networks into node embeddings. We implemented the proposed and two baseline embedding methods on realistic Ethereum network for a predictive tasks with practical relevance, namely phishing/non-phishing node classification. Experimental results demonstrated the effectiveness of the proposed T-EDGE embedding method, meanwhile indicating that a temporal weighted multidigraph can more comprehensively represent the temporal and financial properties of dynamic transaction networks. 
Moreover, this work open up researches of graph embedding in new domain, financial transaction network. Traditional random walk based methods can be extended to a temporal version with temporal walks and edge sampling strategies.
For future work, we can use the proposed embedding method to investigate more applications of Ethereum or extend the current framework to analyze other large-scale temporal or domain-dependent networks.

\section*{Author Contributions}

All authors performed analyses, discussed the results, and contributed to the text of the manuscript.

\section*{Funding}
The work described in this paper was supported by the National Key Research and Development Program (2016YFB1000101), the National Natural Science Foundation of China (61973325, 61503420)  and the Fundamental Research Funds for the Central Universities under Grant No.17lgpy120.

\section*{Acknowledgments}
This manuscript has been released as a pre-print at \url{https://arXiv.org} (Wu \textit{et al.}~\cite{wu2019tedge}).


\section*{Data Availability Statement}

The datasets presented in this study can be found in online repositories (\small{\url{https://github.com/lindan113/xblock-network_analysis/tree/master/Phishing\%20node\%20classification}}), and is also accessible on our dataset website XBlock (\url{http://xblock.pro}).

\bibliographystyle{frontiersinSCNS_ENG_HUMS} 
\bibliography{Frontiers_tedge_submit_arxiv}


\end{document}